\documentclass[copyright]{eptcs}
\providecommand{\event}{WWV 2012}
\usepackage{breakurl}

\usepackage[utf8]{inputenc} 

\usepackage{graphicx} 

\usepackage{booktabs} 
\usepackage{array}    
\usepackage{paralist} 
\usepackage{verbatim} 
\usepackage{subfig}   
\usepackage{listings}
\usepackage{xspace}

\usepackage{fancyhdr} 
\lstnewenvironment{lstxsd}[1][]{%
  \lstset{language=xml,frame=single,#1}%
}{}
\lstnewenvironment{lstjava}[1][]{%
  \lstset{language={},numbers=left,numberstyle=\tiny,numbersep=5pt,
          xleftmargin=\parindent,xrightmargin=\parindent,frame=single,#1}%
}{}
\lstnewenvironment{lstline}[1][]{%
  \lstset{language={},xleftmargin=\parindent,basicstyle=\ttfamily,#1}%
}{}
\lstnewenvironment{lsterlang}[1][]{%
  \lstset{language={},numbers=left,numberstyle=\tiny,numbersep=5pt,
          xleftmargin=\parindent,xrightmargin=\parindent,#1}%
}{}
\lstnewenvironment{lstoutput}[1][]{%
  \lstset{language=erlang,frame=trBL,frameround=tttt,
          basicstyle=\footnotesize\ttfamily,
          xleftmargin=2em,xrightmargin=2em,#1}%
}{}

\newcommand{\LET}{\texttt{?LET}\xspace}

\title{Automatic WSDL-guided Test Case Generation\\
       for PropEr Testing of Web Services}
\author{Leonidas Lampropoulos $^{1}$ \hspace*{2em} Konstantinos Sagonas $^{2,1}$
\institute{%
  $^{1}$ School of Electrical and Computer Engineering,
  National Technical University of Athens, Greece\\
  $^{2}$ Department of Information Technology, Uppsala University, Sweden
}
\email{$\{\,$leolamp$,\,$kostis$\,\}\,$@softlab.ntua.gr \hspace*{2em} kostis@it.uu.se}
}
\def\titlerunning{Automatic WSDL-guided Test Case Generation for PropEr Testing of Web Services}
\def\authorrunning{L. Lampropoulos and K. Sagonas}

\begin{document}
\maketitle

\begin{abstract}
With web services already being key ingredients of modern web systems,
automatic and easy-to-use but at the same time powerful and expressive
testing frameworks for web services are increasingly important. Our
work aims at fully automatic testing of web services: ideally the user
only specifies properties that the web service is expected to satisfy,
in the form of input-output relations, and the system handles all the
rest. In this paper we present in detail the component which lies at
the heart of this system: how the WSDL specification of a web service
is used to automatically create test case generators that can be fed
to PropEr, a property-based testing tool, to create structurally valid
random test cases for its operations and check its responses. Although
the process is fully automatic, our tool optionally allows the user to
easily modify its output to either add semantic information to the
generators or write properties that test for more involved
functionality of the web services.
\end{abstract}

\section{Introduction} \label{sec:intro}


Web services are an essential part of modern web systems, especially
since the appearance of the Service-Oriented Architecture (SOA).
Testing web services, however, is an extremely slow and painful
process, mainly due to the overly verbose nature of XML SOAP messages
which makes writing test cases by hand not a practical option. Many of
the existing tools for testing web services help speed up this
process, up to a point, but when the web service functionality becomes
involved they fail to assist the tester in testing the web services
in an easy and straightforward manner.


One approach that could be used to make the testing of involved web
services easier is property-based testing (PBT). The idea of PBT is to
express the properties that a program must satisfy in the form of
input-output relations, and present the general structure of valid
input messages, while letting the system handle the creation of
progressively more complex test cases in an attempt to find a
counter-example for the property. Property-based testing is gaining
popularity, especially in the community of functional programming
languages where tools such as QuickCheck (for Haskell and Erlang) or
PropEr (for Erlang) exist.
Property-based testing applied to web services shares the same problem
as other testing approaches: the generators would be most cumbersome
to write manually. This is where our tool comes in: It automatically
creates test case generators and simple properties to be tested based
on the WSDL specification of the web service and feeds them to PropEr
for execution.

To give an idea of how one can use our tool, in
Figure~\ref{fig:response_testing} we show a small example of
performing fully-automated response testing on a web service that, at
the time of this writing (May 2012), can be accessed freely on the
web.
In the command shown in the figure, our tool read the WSDL
specification of this web service and, based on the types and
operations that were specified there, created a file with test case
generators and a property called
\texttt{prop\_ChangeCookingUnit\_responds}. After compiling and
loading this file, it executed 100 random tests on the web service to
check its responsiveness and the web service responded with valid SOAP
messages on all these tests. Our tool however, as we will see, is
designed for more involved testing than checking that a web service
responds to random valid requests. Optionally, it allows the user to
easily modify the created generators and properties in order to test
more involved properties that the web service must satisfy.

\begin{figure}
\begin{lstoutput}
Eshell V5.9 (abort with ^G)
1> proper_ws:response_check("http://www.webservicex.net/ConvertCooking.asmx?WSDL").
Testing property: prop_ChangeCookingUnit_responds
..... (100 dots) .....
OK: Passed 100 test(s).
true
\end{lstoutput}
\caption{Automatic response testing of a web service in an Erlang
  shell using our tool.}
\label{fig:response_testing}
\end{figure}


In this paper we focus on the description of the methods that our tool
employs to handle this automatic test case generation. Similar ideas
have already appeared in a number of papers, however the corresponding
tools have yet to mature, while our design can lead to much deeper and
more thorough testing possibilities because of its integration with
PropEr, a state-of-the-art property based testing tool.


The rest of the paper is organized as follows.
The next section presents the system architecture of our tool and
reviews its key components: PropEr, Yaws, and xmerl.
Section~\ref{sec:automatic} describes the techniques used to handle
automatic creation of test case generators and properties, along with
some examples to show the form of the output code.
In Section~\ref{sec:response_testing}, we show how to use our tool to
test a single operation of a web service, fully automatically.
Section~\ref{sec:related} compares our work with related research and
tools already available, while the last section draws conclusions and
presents ideas for future work.

\section{System Architecture and its Components} \label{sec:architecture}

\begin{figure}[!b]
\centering
\includegraphics[height=.245\textheight]{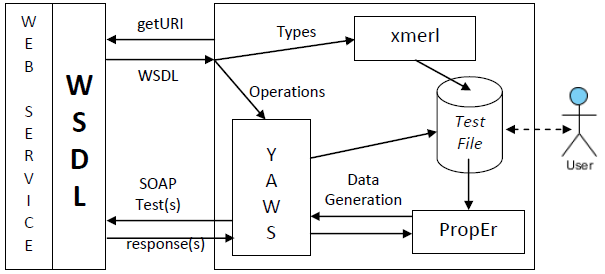}
\caption{System architecture of our property-based testing framework.}
\label{fig:architecture}
\end{figure}

Figure~\ref{fig:architecture} shows the architecture of our testing
framework. Given a URI, the testing starts by obtaining the WSDL
specification of the web service. This specification is then fed into
two different Erlang tools, Yaws and xmerl, which will be briefly
described later in this section. Using xmerl we extract all the type
information associated with the WSDL specification, while using Yaws
we extract needed information for all supported (SOAP) operations.
These two pieces of information are then used to create a testing file
(\texttt{proper\_ws\_autogen}) with Erlang code that contains PropEr
generators and properties ready for use. Then, the user can
(optionally) modify this auto-generated file to add his own properties
or refine the generators. The testing file is then given as input to
PropEr, which generates random test cases, invokes the web service
(using Yaws as a SOAP wrapper) and then analyzes the result.

\paragraph{WSDL}

WSDL is the leading specification for web services in XML
format~\cite{wsdl_spec}, describing the web service in full:
operations, input and output message types, locations, bindings, port
types, etc.

Every WSDL specification contains (or references) an XSD schema
inside, describing the types of the messages needed to invoke the web
service~\cite{xsd_structure_spec,xsd_datatypes_spec}. The types of
these messages are divided in two large categories: simple and complex
types. Simple types can be either primitive data types, such as
floats, integers, strings, etc., aggregates of the above, such as
lists and unions, or restricted versions of them, like enumerations or
range-constrained integers. Complex types on the other hand are
derived (extended or restricted) based on other types, which in turn
are either simple or complex. Usually, complex types are created by
forming element aggregates - sequences or choices.

In addition to types, WSDL also describes the operations the web
service provides, along with information linking the input and output
messages of an operation with types defined in the XSD schema.

\paragraph{PropEr}

Property-based testing is a relatively new approach to software
testing. The user specifies the general structure of valid inputs,
together with properties expected to hold about the input-output
relation. PropEr~\cite{proper_tool} is an Erlang-based PBT tool that
can receive this information and create progressively more complex
test cases, execute them and then monitor the response to make sure it
conforms with the specified properties. In addition, should a failing
test case occur, PropEr will try to locate the part of the test case
that is actually responsible for the fault by \emph{shrinking} (i.e.
simplifying) the offending test case. Since properties are written in
the host language (Erlang), the user can utilize all of its
expressivity to successfully describe a vast range of input-output
relations.

To use PropEr for web service testing, we first need to specify the structure
of the SOAP messages that will be used to invoke the web service operations. 
Then we need a property that will receive the web service's response and use 
it for validation. Both of these, are handled by our tool. By parsing the 
WSDL specification we create valid PropEr generators, while a sample property
is created, that tests that no exceptions (SOAP Faults) happened and that the 
XML response was well formed.

\paragraph{Yaws}

Yaws is one of the most widely used Erlang HTTP web servers~\cite{yaws}.
Yaws uses an XML parser called Erlsom to handle the encoding and decoding
of SOAP messages, a parser module faster and more user friendly than
the xmerl module of the Erlang distribution, imposing however a few
additional limitations. In our framework, Yaws is used at two
different times: in the beginning, in order to extract all the
supported SOAP operations from the WSDL specification, and during the
actual testing phase, as an intermediary between PropEr and the web
service, wrapping the data generated by PropEr in a valid SOAP
structure, invoking a web service operation with the formed SOAP
message, retrieving the result and returning it in the form of an
Erlang tuple to PropEr for further analysis.

\paragraph{xmerl}

Xmerl is an XML parser, included in the Erlang/OTP distribution~\cite{xmerl}.
It transforms any XML to a (rather verbose) Erlang structure containing all the 
information contained in the original XML document. In our framework, xmerl is 
used to parse the XSD Schema of the WSDL specification into an Erlang structure, in 
order to extract afterward the typing information needed to create PropEr generators.

\section{Automatic Creation of Generators and Properties} \label{sec:automatic}

\subsection{Automatic Creation of Test Case Generators from WSDL Types}

At the heart of our tool lies the automatic creation of PropEr generators from 
the types described in the WSDL specification. To that end, we introduce an 
intermediate Erlang representation of the WSDL types; a representation that can 
be directly mapped to PropEr generators, while at the same time is
easier to work with and can be used to handle all constraining facets
of the XSD Schema.

Using an XSD schema, one can describe increasingly complex types by combining
smaller, simple data types into complex ones. Therefore, the first step towards 
creating the intermediate representation is mapping the primitive data types 
into Erlang tuples. The following table shows this mapping for some of the
most used basic types:

\begin{center}\footnotesize
  \begin{tabular}{cc}
    \toprule
    Simple Type & Erlang Intermediate Representation \\
    \midrule
    boolean & \texttt{$\{$erlsom\_bool, bool, []$\}$}\\
    float & \texttt{$\{$erlsom\_string, float, $\{$inf, inf$\}\}$}\\
    double & \texttt{$\{$erlsom\_string, float, $\{$inf, inf$\}\}$}\\
    integer & \texttt{$\{$erlsom\_string, integer, $\{$inf, inf$\}\}$}\\
    nonPositiveInteger & \texttt{$\{$erlsom\_string, integer, $\{$inf, 0$\}\}$}\\
    negativeInteger & \texttt{$\{$erlsom\_string, integer, $\{$inf, -1$\}\}$}\\
    long & \texttt{$\{$erlsom\_string, integer, $\{$-1 bsl 63, 1 bsl 63 -1$\}\}$}\\
    int & \texttt{$\{$erlsom\_int, integer, $\{$-1 bsl 31, 1 bsl 31 - 1$\}\}$}\\
    short & \texttt{$\{$erlsom\_string, integer, $\{$-1 bsl 15, 1 bsl 15 - 1$\}\}$}\\
    byte & \texttt{$\{$erlsom\_string, integer, $\{$-1 bsl 7, 1 bsl 7 - 1$\}\}$}\\
    nonNegativeInteger & \texttt{$\{$erlsom\_string, integer, $\{$0, inf$\}\}$}\\
    positiveInteger & \texttt{$\{$erlsom\_string, integer, $\{$1, inf$\}\}$}\\
    unsignedLong & \texttt{$\{$erlsom\_string, integer, $\{$0, 1 bsl 64 - 1$\}\}$}\\
    unsignedInt & \texttt{$\{$erlsom\_string, integer, $\{$0, 1 bsl 32 - 1$\}\}$}\\
    unsignedShort & \texttt{$\{$erlsom\_string, integer, $\{$0, 1 bsl 16 - 1$\}\}$}\\
    unsignedByte & \texttt{$\{$erlsom\_string, integer, $\{$0, 1 bsl 8 - 1$\}\}$}\\
    string & \texttt{$\{$list, $\{\{$range, 0, inf$\}$, $\{$erlsom\_int, integer, $\{$32, 127$\}\}\}\}$}\\
    \bottomrule
  \end{tabular}
\end{center}

The above format allows us to handle most facets --- like the min/max facets ---
simply by altering the related values according to the schema. In addition, the
same list tuple is used to combine elements inside complex structures, when
their \texttt{(minOccurs,maxOccurs)} attributes are different than~\texttt{(1,1)}.

Another important issue that needs addressing is how to combine these
simple data types into complex ones. In an XSD schema, there are three
main combinators for this: all, sequence and choice. The following
table shows how to create a complex type. We assume we have the name
of the complex type in the \texttt{TypeName} variable, and the
generators of the ``elements'' of the complex type in the
\texttt{Generators} variable.

\begin{center}\footnotesize
  \begin{tabular}{cc}
    \toprule
    Indicator & Erlang Intermediate Representation \\
    \midrule
    all & \texttt{$\{$TypeName, $\{$tuple, Generators$\}\}$}\\
    sequence & \texttt{$\{$TypeName, $\{$tuple, Generators$\}\}$}\\
    choice & \texttt{$\{$TypeName, $\{$union, Generators$\}\}$}\\
    \bottomrule
  \end{tabular}
\end{center}

This intermediate representation can be easily mapped to Erlang code.
In the following table we see how most of the tuples of the
intermediate representation representing simple types are mapped to
code:
\begin{center}\footnotesize
  \begin{tabular}{cc}
    \toprule
    Erlang Intermediate Representation & Code\\
    \midrule
    \texttt{$\{$\_, integer, $\{$inf, inf$\}\}$} & \texttt{integer()}\\
    \texttt{$\{$\_, integer, $\{$Min, Max$\}\}$} & \texttt{integer(Min, Max)}\\
    \texttt{$\{$\_, float, $\{$inf, inf$\}\}$} & \texttt{float()}\\
    \texttt{$\{$\_, float, $\{$Min, Max$\}\}$} & \texttt{float(Min, Max)}\\
    \texttt{$\{$\_, bool, \_$\}$} & \texttt{union([true, false])}\\
    \bottomrule
  \end{tabular}
\end{center}

The first argument is one of \texttt{erlsom\_int},
\texttt{erlsom\_string}, or \texttt{erlsom\_bool}. This argument is
used afterward to wrap the resulting code inside a \LET macro if
needed like this:

\begin{center}
  \texttt{?LET(Gen, Code, \%TYPE\%\_to\_list(Gen))}
\end{center}
which converts the generated instance to a string, if Yaws expects it as such.

\subsection{Example}

The following XSD Schema is an example of a type specification of a web service 
which we will use to show how generators are created.

\begin{lstxsd}
  <complexType name="ProductType">
    <sequence>
      <element maxOccurs="1" minOccurs="1" name="name" type="xsd:string"/>
      <element maxOccurs="1" minOccurs="1" name="price" type="xsd:positiveInteger"/>
      <element maxOccurs="1" minOccurs="1" name="shipInfo" type="impl:ShipInfo"/>
    </sequence>
  </complexType>
  <simpleType name="PaymentType">
    <restriction base="xsd:string">
      <enumeration value="visa"/>
      <enumeration value="paypal"/>
      <enumeration value="deposit"/>
    </restriction>
  </simpleType>
  <complexType name="ShipInfo">
    <sequence>
      <element maxOccurs="1" minOccurs="1" name="paymentInfo" type="impl:PaymentType"/>
      <element maxOccurs="1" minOccurs="1" name="address" type="xsd:string"/>
    </sequence>
  </complexType>
  <element name="Order">
    <complexType>
      <sequence>
        <element maxOccurs="unbounded" minOccurs="1" name="products" type="impl:ProductType"/>
      </sequence>
    </complexType>
  </element>
  <element name="Product" type="impl:ProductType"/>
\end{lstxsd}


Suppose we want to create a generator for an operation
\texttt{"placeOrder"} of a web service that takes as an argument a
single element \texttt{"Order"}. In the front end of the
implementation, we use the information provided by the xmerl module
after parsing the schema to go through all elements and their types
recursively, passing through all the nodes of the XSD schema in a
depth-first search style. In every such node, we create a tuple that
contains both name and type information for the node and all its
successors.

Simple types yield tuples that contain only type information, along
with an atom to facilitate compatibility with Erlsom and Yaws. In our
example, we encounter three different simple types. Two of them,
namely \texttt{xsd:string} and \texttt{xsd:positiveInteger} represent
primitive data types of the WSDL specification, whereas
\texttt{tns:PaymentType} is a user defined simple type: a restriction
upon the string type. The tuples associated with these types in our
tool are:
\begin{lstline}
xsd:positiveInteger : {erlsom_string, integer, {1, inf}}
\end{lstline}
The \texttt{positiveInteger} primitive data type is mapped by default to the
above tuple.  Its first element, the Erlang atom \texttt{erlsom\_string}, states
that for compatibility with Yaws it must be converted to String before wrapped
in a valid SOAP structure, its second element states that the base PropEr
generator is \texttt{integer()}, and the last denotes the range $[1,+\infty)$. 
\begin{lstline}
xsd:string : {list, {{range, 0, inf}, {erlsom_int, integer, {32, 127}}}}
\end{lstline}
This simple type shows how strings are mapped to intermediate tuples.
The first element shows that the entire tuple represents a list and
the second argument specifies the length of the list (in this case it
is arbitrary) and the inner type (its intermediate representation) of
the list (in this case a character; since strings in Erlang are lists
of numbers, the generator creates such lists with integers in the
range 32--127, printable characters in the ASCII table).

\begin{lstline}
tns:PaymentType: {elements, ["visa", "paypal", "deposit"]}
\end{lstline}
The final simple type is a user defined restriction upon the basic
string data type, defining an enumeration of the acceptable values. As
a result the only acceptable values are \texttt{"visa"},
\texttt{"paypal"} or \texttt{"deposit"} and the intermediate tuple
denotes just that.

Elements yield tuples that contain type and naming information. The
naming information contains the name of the element (if it exists) as
well as namespace information of the XSD Schema. For example the
\texttt{"price"} element:
\begin{lstline}
<element maxOccurs="1" minOccurs="1" name="price" type="xsd:positiveInteger"/>
\end{lstline}
is represented with the tuple:
\begin{lstline}
{{price, ['ProductType'], 'http://bar'}, {erlsom_string, integer, {1, inf}}}
\end{lstline}

Finally, complex types usually define aggregates of simple types. In
our example, we encounter three complex types, all of which combine
child elements with a \texttt{sequence} combinator - other options
would include \texttt{all} or \texttt{choice}. For example, the
\texttt{"ShipInfo"} complex type is represented as:

\begin{lstline}
{{'ShipInfo',[],'http://bar'},
  {tuple,
    [{{paymentInfo,['ShipInfo'],'http://bar'},
      {{'PaymentType',[],'http://bar'},
       {elements,["visa","paypal","deposit"]}}},
     {{address,['ShipInfo'],'http://bar'},
      {list, {{range,0,inf}, {erlsom_int,integer,{32,127}}}}}]}}
\end{lstline}

This shows that besides the name information, the intermediate
representation consists of a \texttt{tuple} atom, which is indicative
of the \texttt{sequence} combinator, and a list of the child
intermediate tuples. In this case, since the \texttt{minOccurs} and
\texttt{maxOccurs} attributes for the child elements are both equal to
one, we include the child generator as is. In case these attributes
implied a collection (e.g., \texttt{unbounded}), we would need to wrap
the inner generators inside another list generator.


The mapping of the intermediate tuples to PropEr generators is pretty
straightforward by design. Special care is only needed to ensure names
for all generators are unique. To that end we use the full path up to
the current node as prefix, leading to rather long but descriptive and
unique names. The generators that are automatically produced for the
previous example by our tool are shown in Figure~\ref{fig:generators}.

\begin{figure}
\begin{lsterlang}[xleftmargin=2em,numbers=left,numbersep=5pt]
generate_Order_1_products_ProductType_name() -> 
  list(integer(32, 127)).

generate_Order_1_products_ProductType_price() ->  
  ?LET(Gen, integer(1, inf), integer_to_list(Gen)).

generate_Order_1_products_ProductType_shipInfo_ShipInfo_paymentInfo_PaymentType() -> 
  elements(["visa", "paypal", "deposit"]).

generate_Order_1_products_ProductType_shipInfo_ShipInfo_address() -> 
  list(integer(32, 127)).

generate_Order_1_products_ProductType_shipInfo_ShipInfo() -> 
  ?LET({Pr_Order_1_products_ProductType_shipInfo_ShipInfo_paymentInfo_PaymentType,
         Pr_Order_1_products_ProductType_shipInfo_ShipInfo_address},
        {generate_Order_1_products_ProductType_shipInfo_ShipInfo_paymentInfo_PaymentType(),
         generate_Order_1_products_ProductType_shipInfo_ShipInfo_address()},
        [Pr_Order_1_products_ProductType_shipInfo_ShipInfo_paymentInfo_PaymentType,
         Pr_Order_1_products_ProductType_shipInfo_ShipInfo_address]).

generate_Order_1_products_ProductType() -> 
  ?LET({Pr_Order_1_products_ProductType_name, 
         Pr_Order_1_products_ProductType_price, 
         Pr_Order_1_products_ProductType_shipInfo_ShipInfo},
        {generate_Order_1_products_ProductType_name(), 
         generate_Order_1_products_ProductType_price(), 
         generate_Order_1_products_ProductType_shipInfo_ShipInfo()},
        [Pr_Order_1_products_ProductType_name, 
         Pr_Order_1_products_ProductType_price, 
         Pr_Order_1_products_ProductType_shipInfo_ShipInfo]).

generate_Order_1_products() -> 
  ?LET(Len, range(1, inf), vector(Len, generate_Order_1_products_ProductType())).

generate_Order_1() -> 
  ?LET(Pr_Order_1_products, generate_Order_1_products(), [Pr_Order_1_products]).
\end{lsterlang}
\caption{Generators which are automatically generated for the XSD
  schema of our example}
\label{fig:generators}
\end{figure}

Its code reveals the generation method. Simple types, like the
\texttt{"PaymentType"}, are mapped almost directly with special care
taken if needed to add a wrapper that converts the generated data to
string if that is required by Yaws. Complex types on the other hand,
need to be represented as lists for Yaws to accept them, and we use
\LET macros to that end.

PropEr would allow for simplicity to create less verbose versions of
these generators with a little extra care for unique naming. For
example the last generator could have been written as:

\begin{lsterlang}[firstnumber=35]
generate_Order_1() ->
  [generate_Order_1_products()].
\end{lsterlang}

While in a small example as the above, this would appear preferable,
it actually makes things more difficult for the user, should she want
to modify the generated test data or the generators. The importance of
the \LET macro can be seen in the following example: Suppose a web
service has an operation that takes as input an element with an ISBN
field amongst others (a semantically valid ISBN is a unique identifier
for a book and consists of 10 single digits, the last of which is
determined by the previous nine via a congruence). For such a web
service, our tool would generate code of the following form:
\begin{lsterlang}[numbers=none]
generate_opName() ->
  ?LET({Pr_opName_ISBN, ...},
        {generate_opName_ISBN(), ...},
        [Pr_opName_ISBN, ...]).
\end{lsterlang}
where the \texttt{generate\_opName\_ISBN()} generator would create a
10-member list of single digit integers. With a small function that
changes the first digit so that the ISBN congruence is satisfied (e.g.,
\texttt{ISBNize()}), one could rewrite the generator as:
\begin{lsterlang}[numbers=none]
generate_opName() ->
  ?LET({Pr_opName_ISBN, ...},
        {generate_opName_ISBN(), ...},
        [ISBNize(Pr_opName_ISBN), ...]).
\end{lsterlang}
This is only possible with the use of the \LET macro that allows a user to
modify and combine the generated data. 

\subsection{Automatic Creation of Properties from WSDL Operations}

In addition to automatically creating generators, our tool also
creates a couple of functions for each web service operation, that
invokes this operation using Yaws. Also, it creates a small property
to test that the web service always responds for random test cases, with a
well formed SOAP message. This approach allows us to find most basic
errors in a web service implementation identifying SOAP:Fault
structures, since SOAP Faults are analogous to Java exceptions.

For the XSD Schema of the previous section, our tool creates the following code:
\begin{lsterlang}[firstnumber=38]
call_placeOrder(Arguments) ->
  inets:start(),
  call_placeOrder(yaws_soap_lib:initModel(?WSDL_URL), Arguments).
    
call_placeOrder(WSDL, Arguments) ->
  yaws_soap_lib:call(WSDL, "placeOrder", Arguments).
\end{lsterlang}
Both functions invoke the \texttt{placeOrder} operation of the
web service feeding \texttt{Arguments} to Yaws to wrap in a SOAP
message. The difference between the two, is that the first function
attempts to start the inets module (an Erlang module that includes an
HTTP client) and also parses the WSDL specification of the web service
in order to create the model Yaws needs in order to make the actual
call. The second function takes the WSDL model of the web service as
an argument and invokes the operation directly, thus being more
efficient than the first.

Finally, the property that is automatically created is the following:
\begin{lsterlang}[firstnumber=46]
prop_placeOrder_responds() ->
  ?FORALL(Args, generate_Order_1(),
           case call_placeOrder(Args) of 
             {ok, _Attribs, [#'soap:Fault'{}]} -> false;
             {ok, _Attribs, _Result_record} -> true;
             _ -> false
           end).
\end{lsterlang}

The responds property is a \texttt{FORALL} macro, creating random data
for the web service using the PropEr generators previously created,
uses the aforementioned calling functions to invoke the operation and
finally pattern matches on the resulting tuple. If Yaws encounters
some problem in parsing the SOAP messages an \texttt{$\{$error,
  Reason$\}$} tuple is returned and the property is not satisfied. Also,
if a \texttt{soap:Fault} struct is encountered as the body of the web
service SOAP response, the web service has failed to respond correctly
and the property is also not satisfied. In any other case, Yaws was
able to parse a valid SOAP answer, so we assume the web service
responded correctly and move on to the next test case. For deeper
result validation the user needs to modify this property (or write
another) utilizing the full power of a property-based testing tool
like PropEr.

For the resulting file to be compilable, we should include all
headers, defines, and imports needed. In addition the tool creates a
function \texttt{answer\_placeOrder} which shows how to extract the
answer record that is returned by Yaws. Finally, the extension also
uses erlsom to output a small \texttt{.hrl} (Erlang header) file that
describes the records used for the responses of the Service. This
header file helps the user in decoding the SOAP message responses
without having to deal with XML parsing.

\section{Response Testing of Web Services} \label{sec:response_testing}

One kind of testing our tool can handle, which is completely
automatic, is response testing. Basically, the output file created
contains a property that invokes the operations of the web service
with random inputs and expects an answer for each invocation. This
basically checks if a web service crashes for a specific input or
similar unwanted behaviors.
Let us see how to use response testing to test an existing web service
that converts between cooking units. This free web service is hosted
at \url{http://www.webservicex.net/ConvertCooking.asmx}. We get its
WSDL specification using the URI
\url{http://www.webservicex.net/ConvertCooking.asmx?WSDL"}.
Its XSD Schema is the following:
\begin{lstxsd}
  <s:element name="ChangeCookingUnit">
    <s:complexType>
      <s:sequence>
        <s:element minOccurs="1" maxOccurs="1" name="CookingValue" type="s:double" />
        <s:element minOccurs="1" maxOccurs="1" name="fromCookingUnit" type="tns:Cookings" />
        <s:element minOccurs="1" maxOccurs="1" name="toCookingUnit" type="tns:Cookings" />
      </s:sequence>
    </s:complexType>
  </s:element>
  <s:simpleType name="Cookings">
    <s:restriction base="s:string">
      <s:enumeration value="drop" />
      <s:enumeration value="dash" />
      ...
      <s:enumeration value="TenCan" />
    </s:restriction>
  </s:simpleType>
  <s:element name="ChangeCookingUnitResponse">
    <s:complexType>
      <s:sequence>
        <s:element minOccurs="1" maxOccurs="1" name="ChangeCookingUnitResult" type="s:double" />
      </s:sequence>
    </s:complexType>
  </s:element>
  <s:element name="double" type="s:double" />
\end{lstxsd}
where the \texttt{Cookings} simple type is a large enumeration with
most of its values omitted in the Schema above. While our tool could
be used to test the above web service fully automatically as shown in
Section~\ref{sec:intro}, the process can also be broken in
intermediate steps to allow for user oversight and/or modifications.
Firstly, we create the output file (by default this file is named
\texttt{proper\_ws\_autogen.erl}), which then can be used directly for
compilation and checking the web service:

\begin{lstoutput}
Eshell V5.9  (abort with ^G)
1> wsdl_handler:generate("http://www.webservicex.net/ConvertCooking.asmx?WSDL").
ok
2> c(proper_ws_autogen).
{ok,proper_ws_autogen}
3> proper:quickcheck(proper_ws_autogen:prop_ChangeCookingUnit_responds()).
..... (100 dots) .....
OK: Passed 100 test(s).
true
\end{lstoutput}

As we can see, the change cooking unit web service was invoked 100
times with random arguments and returned a correctly formed result
each time.

Now for an example of a service that crashes, we created our own web
service using eclipse and tomcat. This example is a variation of a
faulty \texttt{lists:delete/2} function~\cite{PropErTypes@Erlang-11},
which we also used to ensure similar behaviour when testing
traditional programs and web services using PropEr.

We created a simple Java implementation of the web service logic, to
make it directly comparable with the original Erlang implementation:
\begin{lstjava}
public Class Delete {
    private String delete(String in, char c, StringBuffer acc) {
        if (in.equals("")) {
            return acc.toString();
        }
        else if (in.charAt(0) == c) {
            return acc.toString().concat(in.substring(1));
        }
        else {
            return delete(in.substring(1), c, acc.append(in.charAt(0)));
        }
    }
    
    public String delete(String in, String c) {
        return delete(in, c.charAt(0), new StringBuffer(""));
    } 
}
\end{lstjava}

We used this code to implement and publish a simple web service in
tomcat. After using our tool to handle the WSDL specification of this
web service, the output file was the one shown in
Figure~\ref{fig:proper_ws_autogen}:

\begin{figure}
\begin{lsterlang}[xleftmargin=2em,numbers=left,numbersep=5pt]
-module(proper_ws_autogen).

-export([call_delete/1, call_delete/2, answer_delete/1]).

-include_lib("proper/include/proper.hrl").
-include("proper_ws_autogen.hrl").

-define(WSDL_URL, "http://localhost:8080/DeleteProject/services/Delete?WSDL").

generate_delete_1_in() -> 
  list(integer(32, 127)).

generate_delete_1_c() -> 
  list(integer(32, 127)).

generate_delete_1() -> 
  ?LET({Pr_delete_1_in, Pr_delete_1_c},
        {generate_delete_1_in(), generate_delete_1_c()},
        [Pr_delete_1_in, Pr_delete_1_c]).

call_delete(Arguments) ->
  inets:start(),
  call_delete(yaws_soap_lib:initModel(?WSDL_URL), Arguments).
    
call_delete(WSDL, Arguments) ->
  yaws_soap_lib:call(WSDL, "delete", Arguments).

prop_delete_responds() ->
  ?FORALL(Args, generate_delete_1(),
           case call_delete(Args) of 
             {ok, _Attribs, [#soap:Fault()]} -> false; 
             {ok, _Attribs, _Result_record} -> true;
             _ -> false
           end).

answer_delete({ok, _, [Answer_record]}) ->
  Answer_record.
\end{lsterlang}
\caption{Auto-generated test file for the \texttt{lists:delete/2} web service.}
\label{fig:proper_ws_autogen}
\end{figure}

Calling PropEr to quickcheck this property reveals a flaw in our implementation.
\begin{lstoutput}
4> proper:quickcheck(proper_ws_autogen:prop_delete_responds()). 
.!
Failed: After 2 test(s).
[[46],[]]

Shrinking .(1 time(s))
[[],[]]
false
\end{lstoutput}

In our implementation we assume that the string $c$ which is supposed
to contain at its first character the character that should be removed
from the string, is not empty. We have two options of fixing it, either
fix our implementation, or remove this test case from the generator. To
demonstrate how easy it is to change the generators created by the extension
we choose the latter. We change the generator of $c$ to:
\begin{lsterlang}[firstnumber=10]
generate_delete_1_c() ->
  ?LET(Len, range(1, 1), vector(Len, integer(32, 127))).
\end{lsterlang}

We could change its second argument to \texttt{range(1,inf)} or
actually remove the \LET macro, but we would need to be careful not to
actually change the output of the generator to a char instead of a
non-empty char list.

Now only valid test cases are created. Testing the response property again we 
get:
\vfill\eject
\begin{lstoutput}
5> proper:quickcheck(proper_ws_autogen:prop_delete_responds()).
..... (100 dots) .....
OK: Passed 100 test(s).
true
\end{lstoutput}

\section{Related Work} \label{sec:related}

Research in testing web services has seen a lot of growth in the past
few years. A variety of tools has emerged handling disparate aspects
of testing, from functional to integration and regression testing; cf.
a survey on the subject~\cite{SOATesting@springerlink-09}. Our tool
can handle automatic functional testing with structurally valid
test cases created based on the WSDL specification of a web service.
Prior research work using a similar idea includes the work of
Bartolini~\textit{et al.}~\cite{bartolini@ICSOC-08} and of
Ma~\textit{et al.}~\cite{automated@IC-08}. Most existing tools have
expanded on the idea of generating XML messages based on a static
analysis of the WSDL specification. Most of them, however, lack in the
aspect of validating the results of the web service and just present
them to the user for inspection.

Amongst the existing frameworks and tools in the area, there are a few that 
stand out. 
Most notably, SoapUI~\cite{soapUI}, one of the most complete testing frameworks
that can handle semi-automated functional testing, amongst other things. This 
tool however does not automatically generate sample test cases, just aids the
user in doing so.
Other tools, such as WSDLTest~\cite{wsdltest@IEEE06} are similar to ours in 
generation strategy, yet user input happens for every script if modifications 
and assertions (for output validation) are needed. Our tool, creates generators 
that allow random test case creation, while any modification by the user to
refine the generators needs to take place only once and will be valid for all
the SOAP messages generated.
Another category of tools is the one that contains WS-Taxi~\cite{taxi@IC-08}. 
This is one of the first tools to have been created based on the idea
of WSDL-based testing. WS-Taxi was first outlined in
2007~\cite{partition@AST-07}, but as stated in the related papers,
while it provides automatic data generation, the tool lacks a test oracle.
Finally, there is a couple of papers and a tool that use Haskell's
QuickCheck to do automatic test case generation. The idea of
Zhang~\textit{et al.}~\cite{ws_quickcheck} is largely similar to our
own: use a property-based testing tool to create generators that allow
for automatic testing. The tool that spawned from this effort,
monadWS~\cite{monadWS@AST-11}, contains a promising comparison with
SoapUI and SoapTest, yet also does not utilize the power of QuickCheck
for deeper validation of the results returned by the web service.

All in all, what makes our tool stand out from the rest is that it 
was designed for use with a property based testing tools with the 
power of PropEr. Our tool handles automatic test data generation as
efficiently and automatically as many other available tools, yet its
design will allow for faster and powerful testing, using properties 
to automatically validate an arbitrary number of progressively more 
complex test cases.

\section{Concluding Remarks and Future Work}

We demonstrated how our tool can be used to automatically test web services, 
with a different approach than the existing tools on the market. Using our 
tool, we were able to successfully test a large number of web services 
available for free, mostly in webservicex.net. From the 30+ web services we 
attempted to test, all but one proved to respond with valid SOAP for all random 
inputs provided by PropEr. The inability to test the remaining one, resulted 
from the existence of a special (german) character in the WSDL specification.
This character was not handled correctly by Yaws, which exited with an error. 

One course for future work is to perfect test case generation. Our
tool, while still under development, can successfully test all free
web services it was tried against, handling most types and
constraining facets appearing in XSD schemas. However, some types that
are not frequently encountered --- like some date formats or the
pattern constraining facet --- are yet to be handled by our tool and
need to be addressed.

Another idea we are currently working on, is to expand on the functionality 
that is tested automatically. The WSDL specification contains information not
only about the input message types, but also the output ones. Using this
information we could create properties that automatically verify not only that
the web service returns a valid SOAP structure, but that the information inside
is correctly typed.

Our main goal now, however, is to handle property-based testing of web
services using our tool and figure out in which areas our tool could
aid the user even further. Successful property-based testing would be
a huge asset for web service testers, speeding up the time needed to
create test cases and doing the actual testing significantly.

\bibliographystyle{eptcs}
\bibliography{ws}

\end{document}